\documentclass[aps,prb,amssymb,showpacs,twocolumn]{revtex4-1}

\newcommand{\etab}{\mbox{\boldmath $\eta $}}

\newcommand{\Fmath}{\mathcal{F}}

\newcommand{\bq}{{\bf q}}

\newcommand{\br}{{\bf r}}

\newcommand{\bR}{{\bf R}}

\newcommand{\vf}{v_{\rm F}}

\newcommand{\beq}{\begin{equation}}
\newcommand{\beqn}{\begin{eqnarray}}
\newcommand{\eeq}{\end{equation}}
\newcommand{\eeqn}{\end{eqnarray}}
\newcommand{\nn}{\nonumber}

\usepackage{graphics,graphicx,amsmath}

\usepackage{tabularx,float}
\usepackage{epsfig}

\usepackage{subfigure}

\usepackage{bm}
\usepackage{wasysym}

\usepackage{bm}
\usepackage{color}
\usepackage{verbatim}

\begin{document}

\bibliographystyle{apsrev4-1}

\date{\today}

\author{R. Rold\'{a}n$^{1,2}$, M. O. Goerbig$^2$, and J.-N. Fuchs$^2$}
\affiliation{\centerline{$^1$Institute for Molecules and Materials, Radboud University Nijmegen, Heyendaalseweg 135, 6525 AJ Nijmegen, The Netherlands}\\
\centerline{$^2$Laboratoire de Physique des Solides, Univ. Paris-Sud, CNRS, UMR 8502, F-91405 Orsay Cedex, France}}

\title{Theory of Bernstein Modes in Graphene}

\begin{abstract}

We present a theoretical description of Bernstein modes that arise as a result of the coupling between plasmon-like collective excitations (upper-hybrid mode) and inter-Landau-level excitations, in graphene in a perpendicular magnetic field. 
These modes, which are apparent as avoided level crossings in the spectral function obtained in the random-phase approximation, 
are described to great accuracy in a phenomenological model. Bernstein modes, which may be measured in inelastic 
light-scattering experiments or in photo-conductivity spectroscopy, are a manifestation of the Coulomb interaction between 
the electrons and may be used for a high-precision measurement of the upper-hybrid mode at small non-zero wave vectors.

\end{abstract}


\maketitle

\section{Introduction}

Most of the electronic properties of graphene, two-dimensional (2D) graphite, may be understood within the picture 
of non-interacting or weakly-correlated massless Dirac particles. \cite{CG07} A notable exception is the recently 
observed fractional quantum Hall effect, \cite{DA09,BK09} which arises in a strong magnetic field when a Landau level
(LL) is only partially filled and when the Coulomb interaction becomes the relevant energy scale due to a quenched 
kinetic energy, similarly to the usual 2D electron gas in semiconductor heterostructures. Apart from this rather particular
situation, the role of the Coulomb interactions may be quantified with the help of the graphene fine-structure constant 
$\alpha_G\equiv e^2/\hbar \epsilon v_F\simeq 2.2/\epsilon$, in terms of the Fermi velocity $v_F\simeq 10^{6}$ m/s and 
the dielectric constant $\epsilon$ of the surrounding medium. 

Even if the value of $\alpha_G$ indicates that graphene should be a moderately correlated material, experimental 
indications for the role of Coulomb interactions in the absence of a magnetic field are rather sparse. 
Recent angular-resolved photoemission experiments have
revealed features that hint at ``plasmaron'' excitations that may be viewed as quasi-particles dressed 
by interaction-induced plasmon excitations.\cite{bostwick10}  Furthermore, an indirect determination of the screened fine structure constant, defined as $\alpha_G^*(\bq,\omega)=e^2/\hbar\vf \epsilon(\bq,\omega)$, where the dynamic screening due to inter-band processes is encoded into the dielectric function $\epsilon(\bq,\omega)$, has been measured by means of
X-ray scattering in graphite.\cite{abbamonte10} In the $\omega\rightarrow 0$ and long wavelength limit, the value given in Ref. \onlinecite{abbamonte10} for graphene (extracting indirectly the polarization of graphene from measurements on graphite) is $\alpha_G^*\simeq 0.14$, i.e. a value that is roughly 
one order of magnitude smaller than the one mentioned above, for the case of freestanding graphene ($\epsilon=1$).
The investigation of interaction-related 
effects may therefore be viewed as one of the major issues in fundamental research on graphene.

Interaction effects are expected to play a role in graphene exposed
to a strong perpendicular magnetic field in the form of plasmon-like excitations.\cite{IWFB07,S07,BM08,RFG09,RGF10} Most saliently, the relativistic character of the electrons
in graphene gives rise to rather exotic linear magneto-plasmons,\cite{RFG09} in addition to the upper-hybrid mode 
(UHM) which is a mixture
of the cyclotron mode and 
the usual 2D plasmon with its low-energy $\sqrt{q}$ dispersion, in terms of the wave vector $q$. In the context
of the 2D electron gas in semiconductor heterostructures, the UHM is coupled via the Coulomb interaction
to inter-LL excitations, and their hybridization gives rise to avoided level crossings, known as Bernstein modes.
\cite{bernstein} Bernstein modes have been measured in 
photo-conductivity spectroscopy and inelastic light-scattering experiments,\cite{batke85,batke86,bangert96,holland,richards} and these
techniques may in principle also be applied to graphene.

In this paper, we study theoretically the Bernstein modes in graphene. These Bernstein modes become apparent in the spectral function,
which takes into account the Coulomb interaction between the electrons in the random-phase approximation (RPA).
We propose a phenomenological model that captures the relevant features, as the position and size of the anticrossings between the UHM and the inter-LL excitations, that appear in the excitation spectrum of this system. Moreover, our results could be used to analyze future experimental results and to identify the different modes. For this aim, we propose two different experimental setups that could be applied to graphene in order to measure the Bernstein modes as well as the dispersion relation of the UHM in this material. In contrast to 2D electrons in semiconductor heterostructures with a parabolic band dispersion and an equidistant LL spacing, one 
observes a plethora of inter-LL transitions in graphene as a consequence of the $\lambda\sqrt{Bn}$ scaling of the relativistic 
LLs, where $\lambda=+$ or $-$ for the conduction and the valence band, respectively, and the integer $n$ denotes the LL index.
This results in a large number of intersections between the inter-LL transitions and the UHM, such that a measurement of 
the Bernstein modes would in principle allow for a high-precision determination of the dispersion relation of the upper-hybrid
mode. 

The paper is organized as follows. In Sec. \ref{sec:Bernst}, we discuss the Bernstein modes which are visible in the 
RPA spectral function (Sec. \ref{sec:RPA}) and propose a phenomenological model (Sec. \ref{sec:phen}) that allows for a 
quantitative description of the position and the strength of the modes. Section \ref{sec:disc} is devoted to possible experimental
observations of Bernstein modes in graphene, and we present our conclusions in Sec. \ref{sec:conc}.

\section{Theoretical description of Bernstein modes in graphene}
\label{sec:Bernst}

\subsection{Bernstein modes in the RPA spectral function}
\label{sec:RPA}

\begin{figure}[t]
  \centering
{\label{ImPiRPAzoom_delt0_02}
\includegraphics[width=0.31\textwidth]{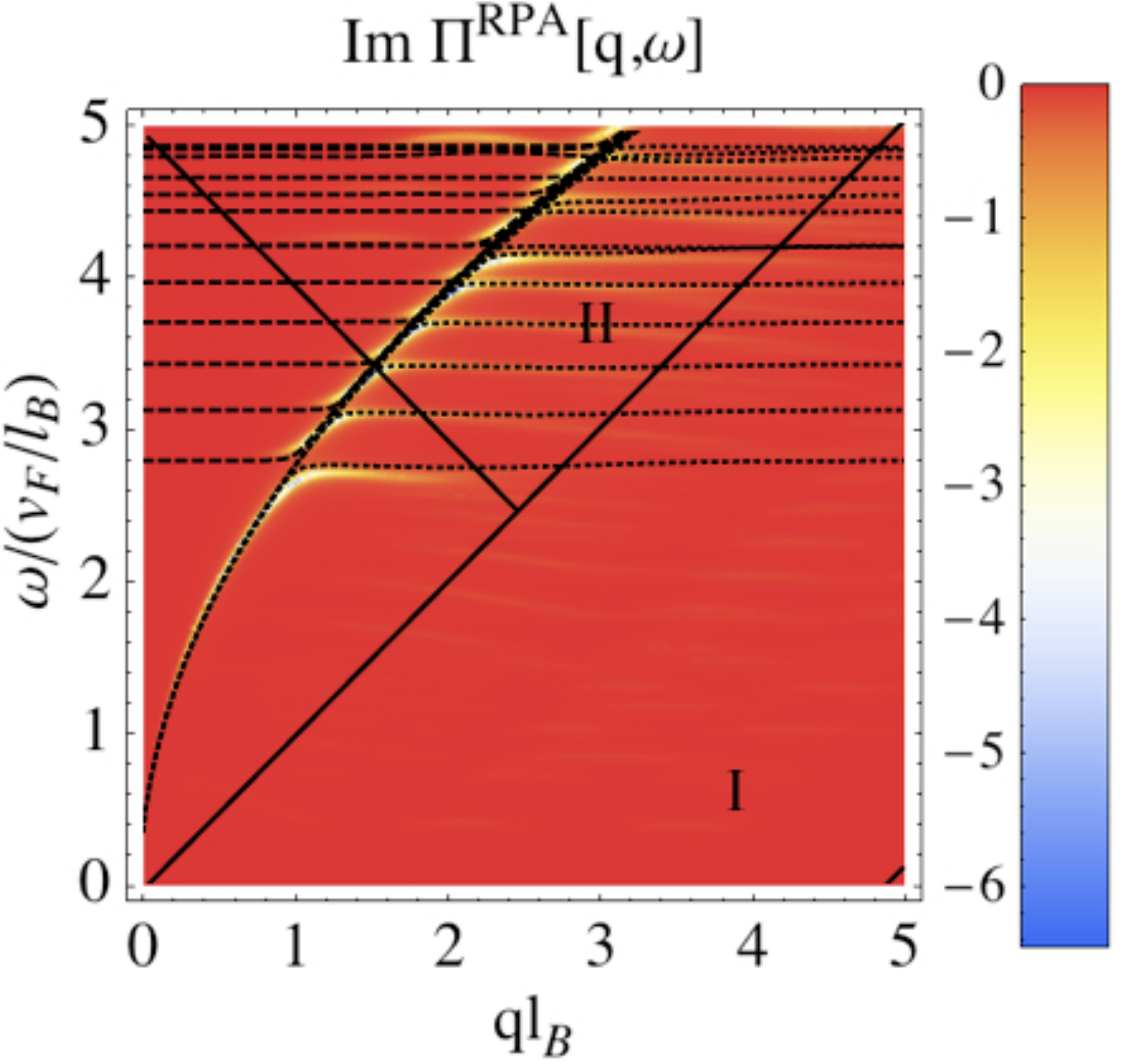}} 
  \caption{(Color online) RPA Spectral function for electrons in graphene, with $N_F=3$, $\delta=0.02v_F/l_B$, 
for a fine-structure constant 
$\alpha_G=1$. The dashed black lines are obtained from the expression (\ref{eq:2modes}) for the coupled modes, for the same parameters
and $\gamma=3/4$. The continuous lines delimit the region for intra-band (I) and that for inter-band (II) excitations.}
  \label{fig01}
\end{figure}

Quite generally, the collective charge excitations generated by the Coulomb interaction between the electrons 
may be obtained from the zeros of the RPA dielectric function 
\beq\label{diel}
\varepsilon_{RPA}(\omega,\bq)=1 - \frac{2\pi e^2}{\epsilon |\bq|}\Pi^0(\omega,\bq),
\eeq
in terms of the polarizability $\Pi^0(\omega,\bq)$ for non-interacting 2D electrons. For electrons in graphene
in the integer quantum Hall regime, the polarizability reads (we use a system of units with $\hbar\equiv 1$) \cite{S07,RFG09}
\beqn\label{eq:LindB}
\nn
\Pi^0(\omega,\bq) &=&\sum_{\lambda n,\lambda'n'}\frac{g[n_F(\lambda n) - n_F(\lambda' n')]}{\omega'(\lambda\sqrt{n}-
\lambda^{\prime} \sqrt{n'})+\omega+i\delta} \\
&&\times 
\frac{|\Fmath_{\lambda n,\lambda'n'}(\bq)|^2}{2\pi l_B^2},
\eeqn
where $n_F(\lambda n)$ is the Fermi distribution function, $\omega'\equiv\sqrt{2}v_F/l_B$ is the characteristic LL frequency, 
$g=4$ represents
the four-fold spin-valley degeneracy, and $\delta$ takes 
into account the disorder-induced LL broadening on a phenomenological level. The form factors $\Fmath_{\lambda n,\lambda'n'}(\bq)$
are due to the graphene-LL wave functions and take into account the chirality of the carriers. They may be expressed in 
terms of the cyclotron variable $\etab=\br-\bR$, where $\br$ is the position of the electron and $\bR$ that of the center
of the cyclotron motion, which is a constant of motion,
\beqn\label{eq:form2}
\nn
\Fmath_{\lambda n,\lambda' n'}(\bq) &=& 1_n^*1_{n'}^*\left\langle n-1\left|
e^{-i\bq\cdot\etab} \right| n'-1\right\rangle \\
&&+\lambda\lambda' 2_n^*2_{n'}^*\left\langle n\left|e^{-i\bq\cdot\etab} \right| n'\right\rangle,
\eeqn
and we have used the short-hand notation $1_n^*\equiv \sqrt{(1-\delta_{n,0})/2}$ and $2_n^*\equiv \sqrt{(1+\delta_{n,0})/2}$.
The states $|n\rangle$ are the usual eigenstates of the harmonic-oscillator operators, $a^{\dagger}a|n\rangle=n|n\rangle$,
where $a=(\eta_x+i\eta_y)/\sqrt{2}l_B$.

The spectral function, i.e. the imaginary part of the RPA polarizability $\Pi^0(\omega,\bq)/\varepsilon_{RPA}(\omega,\bq)$,
is shown in Fig. \ref{fig01}, where we have chosen, for illustration reasons, 
a value of $\alpha_G=1$ for the graphene fine-structure constant. The Fermi
level $E_F=\lambda \omega'\sqrt{N_F}$ is fixed in the LL $N_F=3$ in the conduction band ($\lambda=+$), and we have chosen 
an impurity broadening of $\delta=0.02 v_F/l_B$. 
The main spectral weight is concentrated in the UHM, which disperses as \cite{CQ74}
\beq\label{eq:uhmode}
\omega_{uh}(q,B)=\sqrt{\omega_p^2(q)+\omega_C^2(B)}
\eeq
and which may be viewed as the high-field descendent of the $B=0$ 2D plasmon with the approximate dispersion relation 
\cite{S86,WSSG06,HS07} 
\beq\label{eq:plasmon}
\omega_p(q)\simeq\sqrt{\frac{2e^2 E_F}{\epsilon}q+\gamma
v_F^2q^2}.
\eeq
The first term in this expression yields the usual $\sqrt{q}$ dispersion of the classical plasmon, \cite{stern67} which is clearly
visible in Fig. \ref{fig01}, while the second
one represents a higher-order correction due to quantum effects. 
The small-$q$ expansion of the RPA polarization function yields a negative prefactor for the quantum
corrections, $\gamma= -\alpha_G^2$ \cite{HS07}, in contrast to non-relativistic electrons. It seems, however, that $\gamma$
depends itself on $q$ and $\alpha_G$, namely at larger values of the wave vector where $\gamma$ crosses over to positive values. 
Here, we use $\gamma$ as a fitting parameter to our RPA results in Fig. \ref{fig01}, where the UHM is best described
by the value $\gamma=3/4$ as noted by Shung,\cite{S86} in the shown wave-vector range and for an interaction parameter of $\alpha_G=1$.
The main effect of the magnetic field is the $q=0$ gap, which the plasmon acquires and that is given by the density-dependent
cyclotron gap $\omega_C(B)=eB v_F^2/E_F$ in Eq. (\ref{eq:uhmode}).

It is clearly visible in Fig. \ref{fig01} that, although the main part of the spectral weight is found in the upper-hybrid
mode, this mode is not a continuous line. Instead, one notices a series of avoided level crossings whenever the upper-hybrid 
mode coincides with the energy of a dispersionless inter-LL transition, $\Omega_{\lambda n,n'}=\omega'(\sqrt{n'}-\lambda \sqrt{n})$, 
where an electron 
is promoted from the LL $n$ in the band $\lambda$ to $n'$ in the conduction band (above the Fermi level). These avoided
level crossings are nothing other than the Bernstein modes, which occur at the wave vectors
\beqn\label{eq:crossings}
ql_B &=& \frac{1}{\gamma}\sqrt{2 N_F}\\
\nn 
&&\times 
\left\{\sqrt{\alpha_G^2+\gamma\left[\frac{(\sqrt{n'}-\lambda\sqrt{n})^2}{N_F} - \frac{1}{4 N_F^2}\right]} - \alpha_G
\right\}
\eeqn
in graphene.

\subsection{Phenomenological model}
\label{sec:phen}

In order to describe the coupling between the UHM and the inter-LL excitations 
within a phenomenological model, we treat the UHM as a bosonic excitation, described by the Hamiltonian
$$H_{uh}=\sum_{\bq}\omega_{uh}(q,B)b_{\bq}^{\dagger}b_{\bq}.$$
Here, the boson operator $b_{\bq}^{(\dagger)}$ is proportional to the 
Fourier component $\rho_{uh}(\bq)$ of the density operator that constitutes the plasmon-type mode. This mode is coupled via
the Coulomb interaction 
$$H_{coupl}=\frac{1}{4}\sum_{\bq} \frac{2\pi e^2}{\epsilon |\bq|}\left[\rho(-\bq)\rho_{uh}(\bq) + \rho_{uh}(-\bq)\rho(\bq)\right],$$
to the inter-LL excitations described by the density components
$$\rho(\bq)=\sum_{\lambda n,\lambda' n'}\mathcal{F}_{\lambda n,\lambda'n'}(\bq)\sum_{m,m'}\langle m|e^{-i\bq\cdot\bR}|m'\rangle 
c_{\lambda n,m}^{\dagger}c_{\lambda'n',m'},
$$
in terms of the fermionic electron operators $c_{\lambda n,m}^{(\dagger)}$. This means that the electronic density
is separated into a part that forms the UHM and another one that describes the inter-LL excitations, and formally
one needs to introduce a constraint to avoid double counting of the electronic degrees of freedom. 

The Coulomb coupling between the UHM and the LL excitations yields a renormalization of $\omega_{uh}$
that may be calculated via a Dyson equation for the dressed propagator of the UHM,
similarly to the magneto-phonon resonance discussed in Ref. \onlinecite{GFKF07}. The avoided level crossing 
is then governed by the polarizability (\ref{eq:LindB}), which is dominated by the resonant term with 
$\omega\simeq \omega_{uh}\simeq \Omega_{\lambda n,n'}$, and the  
equation giving the poles of the dressed propagator
reduces to  
\beq\label{eq:Dyson}
[\omega^2 - \omega_{uh}^2][\omega^2 - \Omega_{\lambda n,n'}^2]=\frac{g\mathcal{V}^2}{4}\omega_{uh}\Omega_{\lambda n,n'},
\eeq
where $\mathcal{V}\equiv (e^2/\epsilon ql_B^2)|\Fmath_{\lambda n,n'}(\bq)|^2$ is the effective coupling constant. From Eq. 
(\ref{eq:Dyson}), one obtains the two solutions
\beqn\label{eq:2modes}
\omega_{\pm}^2 &=& \frac{\omega_{uh}(q)^2+\Omega_{\lambda n,n'}^2}{2} %
\\ \nn 
&& \pm \sqrt{\frac{[\omega_{uh}^2-\Omega_{\lambda n,n'}^2]^2}{4}+\frac{g\mathcal{V}^2}{4}\omega_{uh}\Omega_{\lambda n,n'}}.
\eeqn

The solutions are plotted in Fig. \ref{fig01} in the form of dashed black lines, and one notices the good agreement with the 
maxima of spectral weight obtained from Eq. (\ref{eq:LindB}).
At resonance [$\omega_{uh}(q)\simeq \Omega_{\lambda n,n'}$], Eq. (\ref{eq:2modes}) may be linearized,
$$
\frac{\omega_{\pm}}{v_F/l_B}\simeq (\sqrt{2n'}-\lambda\sqrt{2n}) \pm \frac{\delta_{\lambda n,n'}(q)}{2},
$$
in terms of the relative splitting parameter
\beq\label{eq:split}
\delta_{\lambda n,n'}(q)=\frac{\sqrt{g}\mathcal{V}}{v_F/l_B}=\sqrt{g}\frac{\alpha_G}{ql_B}|\Fmath_{\lambda n,n'}(\bq)|^2.
\eeq
One notices in Fig. \ref{fig01}
that the UHM couples more strongly to inter-LL excitations that emerge from  the 
inter-band regime (region II), with $\lambda=-$, than to those associated with intra-band transitions (region I).
Indeed, the relative splitting parameter (\ref{eq:split}) is proportional to 
the $|\Fmath_{\lambda n,n'}(\bq)|^2$, which is important in the regions of the spectrum which correspond to the 
particle-hole continuum in the absence of the magnetic field.\cite{RFG09,RGF10} Because the UHM does not
enter the intra-band region (I) of this continuum, the coupling to the corresponding inter-LL excitations (with $\lambda=+$)
is strongly suppressed, in contrast to those in the inter-band part of the particle-hole continuum, where the upper-hybrid
mode enters.

\section{Discussion and implications for experiments}
\label{sec:disc}

Experimentally, Bernstein modes have been observed in 2D electron systems in semiconductor heterostructures. \cite{batke85,batke86,bangert96,holland,richards} In contrast to three-dimensional metals, \cite{wysmolek06} 
the main difficulty stems from the 
vanishing plasmon dispersion at $q=0$, such that a direct measurement of Bernstein modes in homogeneous 2D systems
via spectroscopic means is impossible. It is therefore necessary to impose a non-zero value of the wave vector
to the system. Two different techniques have been successfully used to do so. The first one consists of a 
grated coupler, with a well-defined periodicity $a$, parallel and in close vicinity to the 2D electron gas. 
\cite{batke85,batke86,bangert96,holland} The wave vector $q_0=2\pi/a$ in the 2D electron gas, at which one may measure the 
electromagnetic response, is thus fixed by the periodicity of the grid via an electromagnetic coupling effect.
This allows one to measure collective charge excitations via transmission spectroscopy (in the far-infrared regime)
at a characteristic (non-zero) value of $q$. 

An alternative technique consists of inelastic light-scattering experiments at a well-defined angle $\theta$
between the incident photon and the vector perpendicular to the 2D electron gas.\cite{richards} In a light-scattering process, 
the photon thus transfers a momentum $q_0=\Delta\omega\sin\theta/c$ to the electrons, where $\Delta\omega=(\omega_i-\omega_s)$ is
the energy difference between the incident and the scattered photon, and $c$ is
the velocity of light. A resonance at the wave vector $q$ and energy $\Delta\omega$ is then interpreted as a 
collective charge excitation, such as a plasmon.

In both techniques, which may in principle also be applied to graphene, 
it turns out to be simpler to study the Bernstein modes at the resonance condition (\ref{eq:crossings})
as a function of the magnetic field while keeping the wave vector fixed. One notices two essential differences
of graphene with respect to a non-relativistic 2D electron gas in semiconductor heterostructures. First,
because the electrons in graphene reside at the surface, the system has the advantage of being directly accessible by surface spectroscopic means, as e.g. scanning tunneling microscopy. 
Second, as a consequence of the non-equidistant LL spacing in graphene, the inter-LL transitions are not multiples of the fundamental cyclotron frequency $\omega_C=eB/m_B$, in terms of the band mass $m_B$ of the semiconducting
host material.  Indeed, in a 2DEG there are, in general, many electron-hole transitions contributing to the same energy, due to the equidistant LLs. However, each particle-hole process contributes to a different energy in graphene
because of the $\sqrt{Bn}$ dispersion of graphene LLs. As a consequence, within the same energy window, there is a higher number of anticrossings between the UHM and the inter-LL transitions (Bernstein modes) in graphene as compared to a 2DEG.\cite{RGF10} 
Because the UHM is determined by the avoided level crossings with inter-LL excitations, this
would, in principle, allow for a more precise measurement of the UHM in graphene as compared to non-relativistic 
electron systems.

\begin{figure}[t]
  \centering
{\label{BernsteinRPA}
\includegraphics[width=0.28\textwidth]{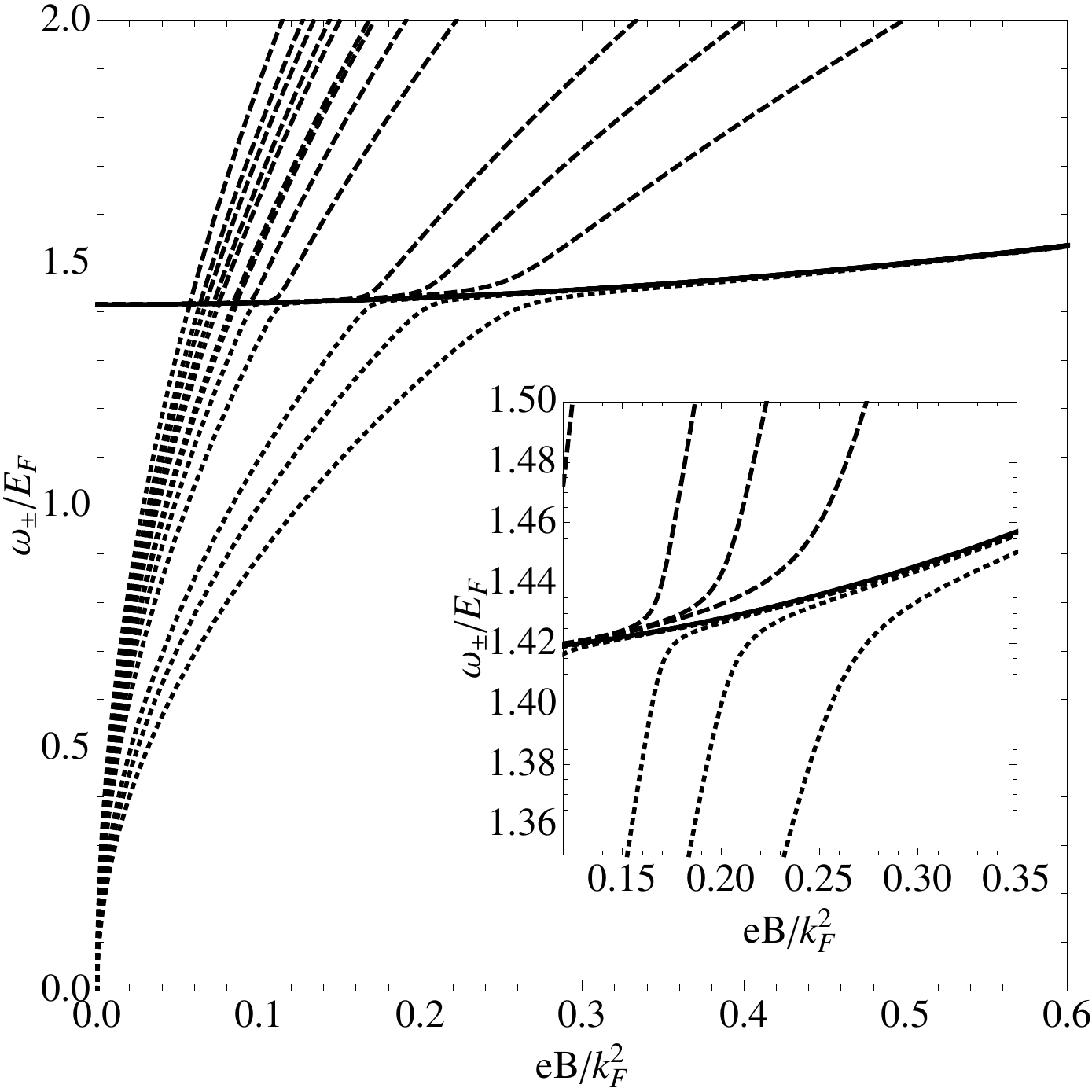}} 
  \caption{Bernstein modes as a function of the magnetic field obtained from Eq. (\ref{eq:2modes}), 
for a fixed value of $q_0/k_F=1$ and $\alpha_G=1$. Dashed lines represent inter-LL excitations, and the continous line
shows the UHM. The inset shows a zoom on the high-field region of the avoided level crossings.}
  \label{fig02}
\end{figure}

In Fig. \ref{fig02}, we have plotted the solutions (\ref{eq:2modes}) of 
the Bernstein modes as a function of the magnetic field instead of the wave vector in order to make a closer 
connection with possible experiments. The wave vector is fixed at $q_0/k_F=1$ and we have chosen again $\alpha_G=1$
for illustration reasons. Notice that, in contrast to the above discussion, the natural units are no longer $v_F/l_B$
for the energy and $l_B^{-1}$ for the wave vector but the density-dependent Fermi energy $E_F=v_Fk_F$ and Fermi
wave vector $k_F=\sqrt{4\pi n_{el}/g}=\sqrt{\pi n_{el}}$. Both units are therefore fixed by the carrier density $n_{el}$,
whereas the index $N_F$ changes as a function of the magnetic field as $N_F\simeq k_F^2l_B^2/2$, 
where we have neglected the discreteness of $N_F$.
The inter-LL excitations now disperse as $\sqrt{B}(\sqrt{n}+\sqrt{n'})$, whereas
the UHM is only weakly dispersing in the plotted magnetic-field range. As one expects from Eq. (\ref{eq:split}),
the Bernstein modes are most prominent at high magnetic fields, and the position of the avoided level crossings are 
easily obtained from Eq. (\ref{eq:crossings}). Neglecting the quantum corrections to the plasmon frequency ($\gamma\rightarrow 0$)
in the small wave-vector limit ($q_0/k_F\lesssim 2\alpha_G /\gamma$), the Bernstein modes are expected at the field values
\beqn\label{eq:crossingsN}
(eB)^{-1} &\simeq& \frac{(\sqrt{n'}-\lambda\sqrt{n})^2}{2\alpha_G q_0k_F} \\
\nn
&& +\sqrt{\frac{(\sqrt{n'}-\lambda\sqrt{n})^4}{4\alpha_G^2 q_0^2k_F^2}
+ \frac{1}{2\sqrt{2}\alpha_Gq_0k_F^3}}.
\eeqn

Notice that the zero-field limit of the dispersion of the UHM, which one may extrapolate from the measurement
of the Bernstein modes, gives direct access to the bare 
graphene fine-structure constant $\alpha_G$. Indeed, if one expresses Eq. 
(\ref{eq:plasmon}) in terms of $k_F$ and $\alpha_G$,
$[\omega_{uh}(q,B\rightarrow 0)/v_F]^2=2\alpha_G k_F q$, one may determine the fine-structure constant in the small-$q$ limit
from the 
expression 
\beq\label{eq:uhB0}
\alpha_G = \frac{1}{2\sqrt{\pi n_{el}}q_0}\frac{\omega_{uh}^2(q_0,B\rightarrow 0)}{v_F^2},
\eeq
However, we emphasize that it is the unscreened graphene fine-structure constant which occurs in these expressions and not
the screened one $\alpha_G^*$. Naturally, the information about $\alpha_G$ is also encoded in the 
size of the mode splitting (\ref{eq:split}), but it seems more delicate to extract because of the complicated $B$-field and $q$ 
dependence of the form factor. 

\section{Conclusions}
\label{sec:conc}

In conclusion, we have theoretically analyzed the Bernstein modes in graphene, which arise as a consequence of the interaction
between plasmon-type modes and inter-LL excitations. The proposed model captures the position and strength of the couplings, as 
compared to the numerical solution of the RPA polarizability. Finally, we have shown that a possible 
measurement of Bernstein modes with the help of techniques that have been successfully used in conventional 2D electron systems
with a parabolic band dispersion \cite{batke85,batke86,bangert96,holland,richards} and that may equally be used in graphene would allow
for a measurement of the dispersion relation of the UHM. 

\acknowledgements 

We acknowledge H. Bouchiat and M. I. Katsnelson for fruitful discussions. This work was funded by ``Triangle de la Physique'',
the EU-India FP-7 collaboration under MONAMI, and the ANR project NANOSIM GRAPHENE under Grant No. ANR-09-NANO-016.

\bibliography{BibliogrGrafeno2}

\end{document}